\documentclass[graybox]{svmult}

\usepackage{type1cm}        
\usepackage{makeidx}         
\usepackage{graphicx}        
\usepackage{multicol}        
\usepackage[bottom]{footmisc}

\usepackage{newtxtext}       %
\usepackage[varvw]{newtxmath}       

\usepackage{subcaption}
\usepackage{cite}

\makeindex             

\begin{document}
	
	\title*{Anomalous Transport and Explicit Symmetry Breaking in Holography}
	\author{Ashis Tamang, Nishal Rai, Karl Landsteiner and Eugenio Meg\'{\i}as}
	
	\institute{
		Ashis Tamang \at Department of Physics, SRM University Sikkim, Upper Tadong, Sikkim, 737102,  India\\
		\email{ashismaniram@gmail.com}
		\and 
		Nishal Rai \at Department of Physics, St. Joseph's College, North Point, Darjeeling, 737104, India \\
		Center for Astrophysics, Gravitation and Cosmology (CAGC), SRM University Sikkim, Upper Tadong, Sikkim, 737102, India \\
		\email{nishalrai10@gmail.com}
		\and
		Karl Landsteiner \at Instituto de F\'{\i}sica Te\'orica UAM/CSIC, C/ Nicol\'as Cabrera 13-15, \\
		Universidad Aut\'onoma de Madrid, Cantoblanco, 28049 Madrid, Spain \\
		\email{karl.landsteiner@csic.es}
		\and 
		Eugenio Meg\'{\i}as \at Departamento de F{\'\i}sica At\'omica, Molecular y Nuclear and \\
		Instituto Carlos I de F{\'\i}sica Te\'orica y Computacional, \\
		Universidad de Granada, Avenida de Fuente Nueva s/n, E-18071 Granada, Spain \\
		\email{emegias@ugr.es}
	}
	%
	%
	\maketitle

	\abstract{We consider a holographic Einstein-Maxwell model in five dimensions with pure gauge and mixed gauge-gravitational Chern-Simons terms to study anomaly-induced transport in the presence of explicit symmetry breaking. We include the full backreaction of the scalar field and gauge fields on the metric and compute the anomalous transport coefficients using Kubo formulae involving charge and energy current correlators. Our findings reveal that, in the presence of explicit symmetry breaking, anomaly-induced transport phenomena can extend beyond anomalous currents and affect the non-anomalous sector as well. The transport coefficients exhibit a clear dependence on the symmetry-breaking mass parameter, highlighting the interplay between quantum anomalies and explicit symmetry breaking in holographic systems. }
	
	\section{Introduction}
	Over the past few decades, the discovery of a deep connection between field theories and gravitational theories has significantly shaped modern theoretical physics. One of the most remarkable developments in this context is the gauge/gravity duality often known as AdS/CFT correspondence which relates certain strongly coupled quantum field theories to classical gravitational theories in higher-dimensional spacetimes. Due to the reason that it relates the theory in higher dimension to some theory with one dimension less it is also called as Holography.  The original and most famous studied example of this duality is the one between ${\mathcal N} = 4$ supersymmetric Yang-Mills (SYM) theory in the large $N$ limit and type IIB string theory in the AdS$_5$ $\times$ S$^5$ space \cite{Maldacena:1997re,Witten:1998qj,Gubser:1998bc}. 
	
	Anomaly induced transport phenomena have become a highly active area of research in relativistic field theories of chiral fermions, extending beyond the reach of conventional perturbative approaches \cite{Kharzeev:2015znc,Bastianelli:2006rx}. In recent years, holographic methods have been extensively employed to investigate these anomalous effects, yielding valuable insights into their underlying mechanisms~\cite{Bhattacharyya:2007vjd,Erdmenger:2008rm,Landsteiner:2011iq,Rai:2023nxe}. A notable example is the study presented in \cite{Megias:2016ery}, where a holographic model with a pure Chern-Simons term was considered. One of the key findings of this work is the identification of a novel phenomenon where anomalies can induce transport effects not only in anomalous currents but also in non-anomalous ones, i.e., those with a vanishing divergence, when symmetries are explicitly broken. In the present work, we aim to extend this analysis to other anomalous coefficients, such as the chiral vortical conductivity. Furthermore, we investigate the role of the mixed gauge-gravitational anomaly in the generation of the non-anomalous currents. To this end, we consider a model with three symmetries, two corresponding to the familiar vector and axial $U(1)$ symmetries of Dirac fermions, with only the axial symmetry being anomalous, and a third symmetry that remains non-anomalous. We also introduce a scalar field that explicitly breaks symmetry and carries charge under both the axial and the non-anomalous symmetries. The boundary value $M$ of this scalar field serves as the parameter controlling the strength of symmetry breaking.
	
	The remainder of this chapter is organized as follows. In Section \ref{model}, we introduce the holographic model and present the corresponding background solutions. Section \ref{fluct} is devoted to the study of linear perturbations and the computation of the relevant correlators. The main results for the anomalous transport coefficients are presented in Section \ref{result}, where we discuss their dependence on the chemical potentials and the symmetry-breaking parameter. Finally, we conclude in Section \ref{disc} with a discussion of the results and  possible  future directions.

	\section{Holographic Model}
	\label{model}
	In this section we introduce the model and study the main properties of the background solution.

        	\subsection{Action}
	\label{sec:action}

        We consider a holographic model that includes both a pure gauge and a mixed gauge-gravitational Chern-Simons term in the action\cite{Tamang:2025ekj,Landsteiner:2013aba,Megias:2016ery}. In order to explore the possibility that the constitutive relation for the non-anomalous current, $\langle J_w^\mu \rangle$, receives anomalous contributions, we extend our model by introducing explicit symmetry breaking via a scalar field. The action of the model is given by
	\begin{eqnarray}
		\mathcal S &=& \dfrac{1}{16 \pi G} \int d^5x\sqrt{-g}
		\Big[
		R + 2\Lambda -\dfrac{1}{4}  F_V^{2} \nonumber -\dfrac{1}{4}  F_A^{2} \nonumber -\dfrac{1}{4}  F_W^{2} \nonumber \nonumber \\&&+ \epsilon^{MNPQR} A_M \left(\dfrac{\kappa}{3}(F_A)_{NP}(F_A)_{QR}+{\kappa}(F_V)_{NP}(F_V)_{QR}+\lambda R^A\,{}_{BNP}R^B\,{}_{AQR} \right) \nonumber \\&&- |D_M \phi|^2 - m^2\phi^2\Big] + \mathcal S_{\rm GHY} + \mathcal S_{\rm CSK}   \,,  
		\label{act}
	\end{eqnarray}
	where $D_M \phi = \left[ \partial_M -i(A_M - W_M) \right] \phi$, and $\Lambda = 6 / \ell^2$  is the cosmological constant with $\ell$ the radius of AdS. The field strength tensors are defined as $(F_S)_{MN}=\partial_M S_N-\partial_N S_M$ for $S = V, A, W$, with $V$ and $A$ denoting the anomalous vector and axial gauge fields, and $W$ representing an extra non-anomalous gauge field. The boundary terms denoted by $\mathcal S_{\rm GHY} $ and $\mathcal S_{\rm CSK}$  correspond to the Gibbons-Hawking-York term and a term induced by the mixed gauge-gravitational anomaly, respectively. The values of the couplings $\kappa$ and $\lambda$ are given in terms of the anomalous coefficients of the field theory as~\cite{Landsteiner:2011iq}
	\begin{equation}
		\label{eq:kappa-lambda_value}
		\frac{1}{16\pi G}\kappa = -\frac{1}{16\pi^2} \,,  \qquad \frac{1}{16\pi G}
		\lambda = -\frac{1}{384\pi^2} \,.
	\end{equation}

	\subsection{Background solution}
	\label{sec:back_sol}
	We will consider the following ansatz for the background metric and the gauge fields
	  \begin{equation}
          \begin{aligned}
		ds^2 &= - \frac{r^2}{\ell^2}  f(r)dt^2 + e^{\chi(r)} \frac{\ell^2}{r^2} \frac{dr^2}{f(r)} + \frac{r^2}{\ell^2} \left( dx^2 + dy^2 + dz^2\right)   \,, \\
		V &= V_t(r)dt \,, \qquad A = A_t(r)dt \,, \qquad W = W_t(r)dt  \,.
        \end{aligned}
	\end{equation}
	The near boundary expansions of fields are given by
	\begin{equation}
		\lim_{r\to\infty} r \cdot \phi(r) = M \,, \quad \lim_{r\to\infty} V_t(r) = \mu_v  \,, \quad \lim_{r\to\infty} A_t(r) = \mu_a  \,, \quad  \lim_{r\to\infty} W_t(r) = \mu_w  \,, \label{eq:nearboundary}
	\end{equation}
	where we have assumed in these expressions that the gauge fields must vanish on the horizon: $S_t(r_h) = 0$ $\; (S= V, A, W)$. For convenience, we introduce the new combinations of gauge fields, $A_\pm \equiv A \pm W$, so that the covariant derivative writes~$D_M\phi = \left[ \partial_M -i(A_-)_M \right] \phi$. With this redefinition, the corresponding chemical potentials take the form $\mu_\pm \equiv \mu_a  \pm \mu_w$. It is also useful to perform the change of variable $u= r_h^2/r^2$, where $r_h$ is the position of the horizon. In the $u$ variable the horizon is at $u_h=1$. In the rest of the manuscript we will set $\ell = 1$.
	
	Using the ansatz introduced above and working with the $u$ variable, the equations of motion for background fields $V$, $A_+$, $A_-$, $\phi$ and $\chi$ are given by
	\begin{eqnarray}
		0 &=& V_t^{\prime\prime} - \frac{\chi'}{2} V_t^\prime \,, \label{eq:Vt} \\
		0 &=&  A_{t+}^{\prime\prime} - \frac{\chi'}{2} A_{t+}' \,, \label{eq:Apt} \\
		0 &=& A_{t-}^{\prime\prime} - \frac{\chi'}{2} A_{t-}' - \frac{e^{\chi}}{ u^2 f} \phi^2 A_{t-}  \,,  \label{eq:Amt}  \\ 
		0 &=&  \phi^{\prime\prime} - \left( \frac{1}{u} + \frac{\chi^\prime}{2} - \frac{f^\prime}{f} \right) \phi^\prime  + \frac{e^{\chi}}{4 u^2 f^2} \left( 3 f + u A_{t-}^2 \right)  \phi  \,,  \label{eq:phi}  \\
		0 &=& \phi^{\prime \, 2} + \frac{3 \chi^\prime}{4u} + \frac{e^{\chi} }{ 4 u f^2} \phi^2   A_{t-}^2   \,, \label{eq:f}  \\
		0 &=& f^\prime - \frac{f}{2u} (4 + u \chi^\prime)  + \frac{e^{\chi}}{2u} ( 4 + \phi^2 ) - \frac{u^2}{6} \left( 2 V_t^{\prime \, 2}  + A_{t-}^{\prime \, 2} + A_{t+}^{\prime \, 2}  \right) \,, \label{eq:chi}
	\end{eqnarray}
	where the primes stand for derivatives with respect to $u$. 
	
\begin{figure}[htbp]
		\centering
		\includegraphics[width=.47\textwidth]{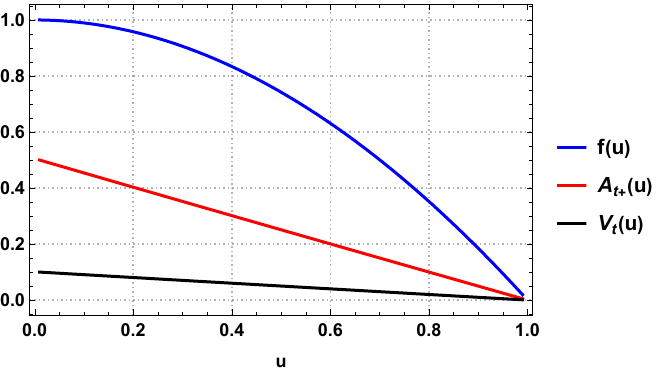}
		\qquad
		\includegraphics[width=.47\textwidth]{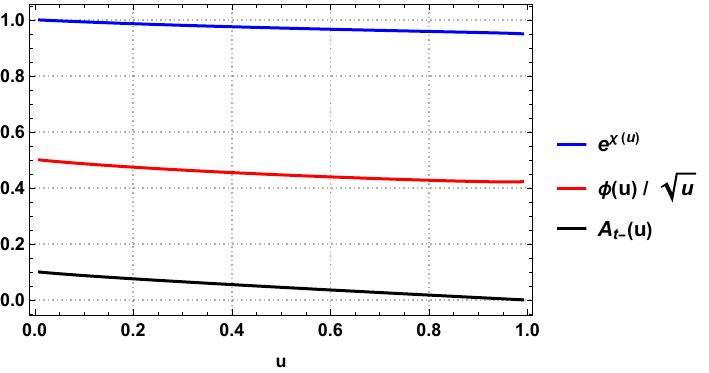}
		\caption{Plot showing the dependence of the background with $u$. Color code: Blue -  $f(u)$, Red - $A_{t+} (u)$,  and Black - $V_t(u)$ (left panel); Blue - $e^{\chi(u)}$, Red - $\phi(u) / \sqrt{u}$, and Black - $A_{t-} (u)$ (right panel). We have set $\mu_v/(\pi T) = 0.1$,  $\mu_a/(\pi T) = 0.3$, $\mu_w/(\pi T) = 0.2$, and $M/(\pi T)=0.5$. \label{fig:i}}
	\end{figure}

	This system of six coupled differential equations can be solved numerically with appropriate boundary conditions given by eq.~(\ref{eq:nearboundary}) together with regularity near the horizon. In order to solve this background we have used the shooting method, shooting from the horizon $u_h=1$ towards the boundary $u=0$. We plot in fig.~\ref{fig:i} the numerical solutions of all the background fields as a function of $u$. 

	\section{Fluctuations}
	\label{fluct}
	We will now consider perturbations of the fields. To study the effect of anomalies, we split the metric and gauge fields into background and a linear perturbation, i.e.   
	\begin{align}
		g_{MN} ={g^{(0)}}_{MN} + \epsilon h_{MN}\,,\,\,\, \qquad   \mathcal S_\mu = S_\mu + \epsilon s_\mu \,, 
	\end{align}
	where $S_\mu$ refers to the background solutions computed in section~\ref{sec:back_sol} for any of the fields $V$, $A_\pm$ (or equivalently $V$, $A$ and $W$), and $s_\mu$ are the corresponding fluctuations $v, a_\pm \equiv a \pm w$ (or equivalently $v$, $a$ and $w$). 
	Then, we will follow the general procedure of Fourier mode decomposition~\cite{Amado:2011zx}
	\begin{eqnarray}
		h_{MN}(u, x^\nu) &=& \int \dfrac{d^4p}{(2\pi)^4} \, h_{MN}(u) \, e^{-i\omega t+i \vec{p} \cdot \vec{x}} \,,\\
		s_{\mu}(u, x^\nu) &=& \int \dfrac{d^4p}{(2\pi)^4} \, s_{\mu}(u) \, e^{-i\omega t+i \vec{p} \cdot \vec{x}}\,.
	\end{eqnarray}
	Without loss of generality, we restrict our analysis to perturbations characterized by momentum $\vec{p}$ directed along the $z$-axis. The conductivities for the dual field theory will be computed by using Kubo formulae involving the retarded correlators of two currents~\cite{Landsteiner:2012kd}. Since we are interested in computing correlators at zero frequency, we can set to zero the frequency-dependent parts in the equations and solve the system up to first order in $p \equiv |\vec{p}|$. To compute the retarded correlators, we need to solve the equations of motion for the fluctuations of the metric and gauge fields. These equations of motion are provided in Appendix A.2 of~\cite{Tamang:2025ekj}. 
	
	To solve the coupled differential equations for the fluctuations we have used the pseudo-spectral method~\cite{trefethen2000spectral, boyd2000spectral}. To get the solutions, we expand the fluctuations as a sum of Chebyshev polynomials in the $u$ direction as
	\begin{equation}
		(a_\pm)_i=\sum_{m=0}^{N-1} (a_\pm^{m})_i T_m (2u-1) \,,  \quad
		v_i=\sum_{m=0}^{N-1}v_i^m T_m (2u-1) \,,  \quad
		h_{it}=\sum_{m=0}^{N-1}h_{it}^m T_m (2u-1) \,,
		\label{cheb}
	\end{equation}
	where $(a_\pm^m)_i $, $v_i^m $ and $h_{it}^m$ are the coefficients of the polynomials in eq.~(\ref{cheb}) with ($i=x,y$). These polynomials are inserted into the equation of motion for the fluctuations, and the collocation points for $u=0$ to $u=1$ are chosen in Gauss-Lobatto grid. By choosing the appropriate number of grid points corresponding to the number of coefficients, one can solve the set of algebraic equations in terms of these coefficients to get the solution for the fluctuations.
	
	\section{Results}
	\label{result}
	
	In this section we will present the results. We present in subsection \ref{sec: result2} the results of the conductivities vs. chemical potentials at finite $M$, and in subsection \ref{sec: result3} the results of the conductivities vs. $M$ for fixed chemical potentials.
	
	\subsection{Results of the conductivities vs. chemical potentials at finite $M$}
	\label{sec: result2}
	
\begin{figure}[htbp]
		\centering
		\includegraphics[width=.4\textwidth]{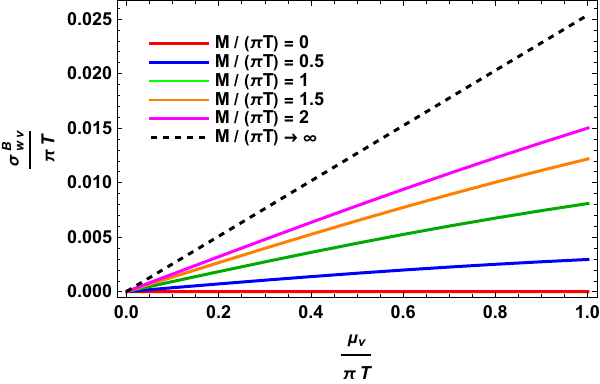}
		\qquad
		\includegraphics[width=.4\textwidth]{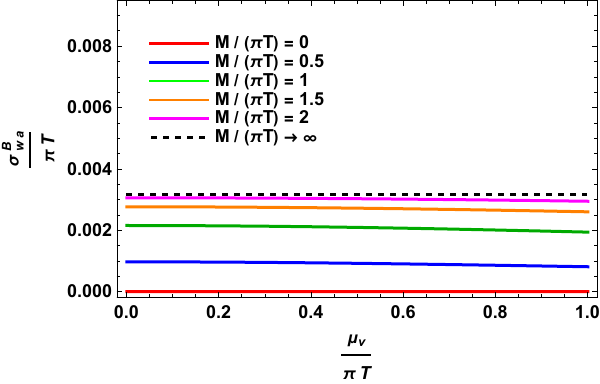}
		\qquad
		\includegraphics[width=.4\textwidth]{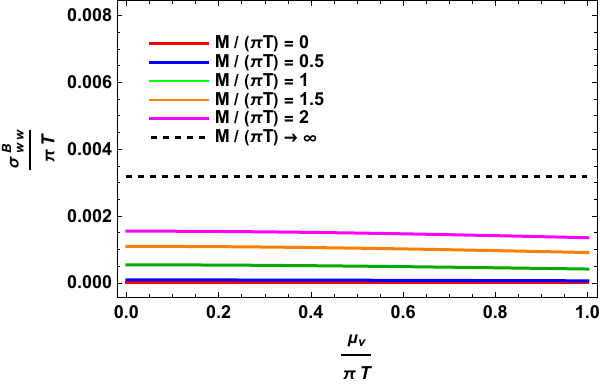}
		\qquad
		\includegraphics[width=.4\textwidth]{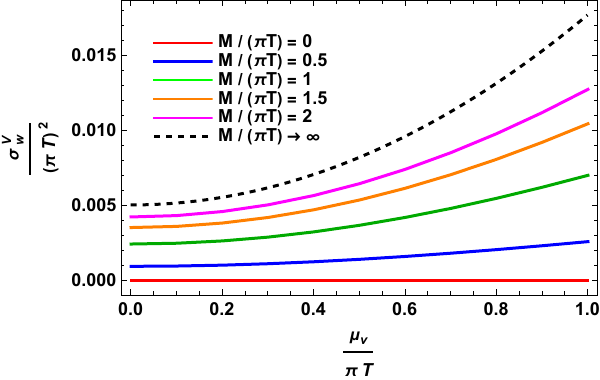}
		\caption{Plots of the conductivities $\sigma^B_{wv}$, $ \sigma^B_{wa}$,  $\sigma^B_{ww}$ and $\sigma^V_{w}$ as a function of to $\mu_v/(\pi T)$. We have set $\mu_a/(\pi T)=0.3$ and $\mu_w/(\pi T)=0.2$.  \label{fig:corywm}}
	\end{figure}

	 We present the results for the conductivities as a function of the chemical potentials, for several values of dimensionless mass parameter $M/(\pi T)$. Here we will present only the $\mu_v$-dependence of the conductivities that induce the non-anomalous current $\vec{J}_w$, i.e. the conductivities appearing~in 
	\begin{equation}
		\vec{J}_w = \sigma^B_{wv} \, \vec{B}_v + \sigma^B_{wa} \, \vec{B}_a + \sigma^B_{ww} \, \vec{B}_w + \sigma^V_{w} \, \vec{\Omega} \,. \label{eq:Jw}
	\end{equation}
        In this expression, $\vec{B}_s$ is the magnetic field associated to symmetry $s \in \{ v, a, w \}$, and~$\vec{\Omega}$ is the vorticity vector. We focus momentarily in this subsection on these conductivities because they allow us to provide an explicit answer to the following question: is it possible to get a {\it chiral magnetic effect} for a non-anomalous symmetry~$w$? We display in fig.~\ref{fig:corywm} the results of the conductivities in eq.~(\ref{eq:Jw}) as a function of $\mu_v / (\pi T)$ in the range $0 \le M/(\pi T) \le 2$. For comparison, we display also the results obtained analytically in the large $M$ limit and which are provided in Appendix C of \cite{Tamang:2025ekj}. 
	
		\subsection{Results of the conductivities vs. $M$ for fixed chemical potentials }
	\label{sec: result3}

\begin{figure}[htbp]
		\centering
		\includegraphics[width=.47\textwidth]{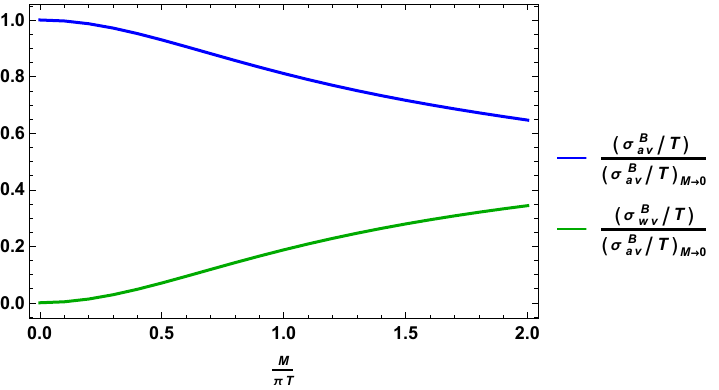}
		\qquad
		\includegraphics[width=.47\textwidth]{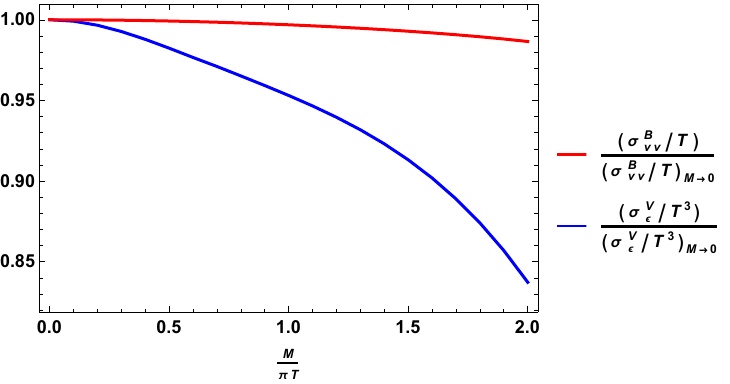}
		\caption{Plots of the chiral magnetic conductivities $\sigma^B_{s v}$ with $s \in \{ a,w \}$ (left panel) and $\sigma^B_{v v}$ along with $\sigma^V_{\varepsilon}$ (right panel), as a function of $M/(\pi T)$. We have set $\mu_v/(\pi T)=0.1$, $\mu_a/(\pi T)=0.3$ and $\mu_w/(\pi T) =0.2$.} \label{fig:comb1}
	\end{figure}

\begin{figure}[htbp]
		\centering
		\includegraphics[width=.47\textwidth]{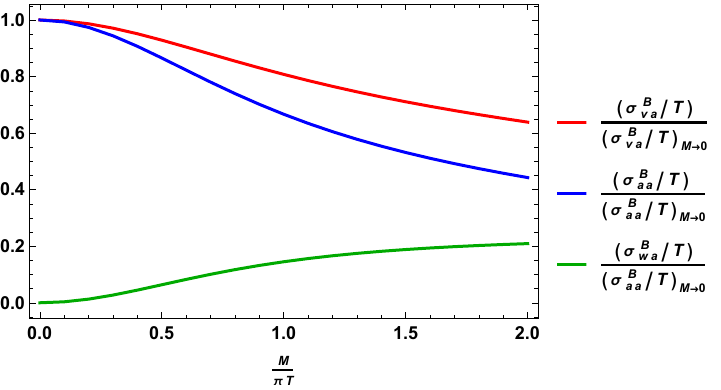}
		\qquad
		\includegraphics[width=.47\textwidth]{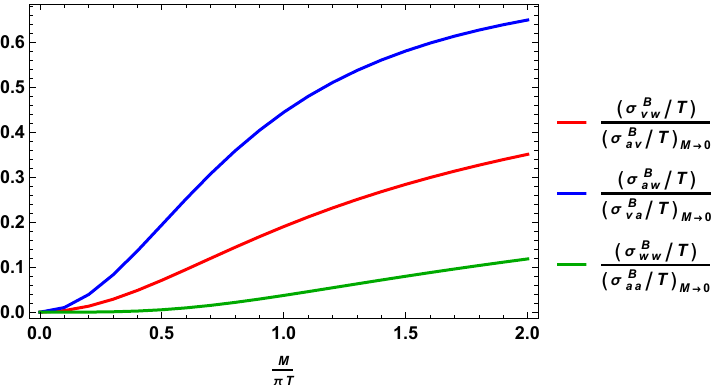}
		\caption{Plots of the chiral magnetic conductivities $\sigma^B_{s a}$ (left panel) and $\sigma^B_{s w}$ (right panel), where $s \in \{ v,a,w \}$, as a function of $M/(\pi T)$. See fig.~\ref{fig:comb1} for further details.} \label{fig:comb2}
	\end{figure}

\begin{figure}[htbp]
		\centering
		\includegraphics[width=.47\textwidth]{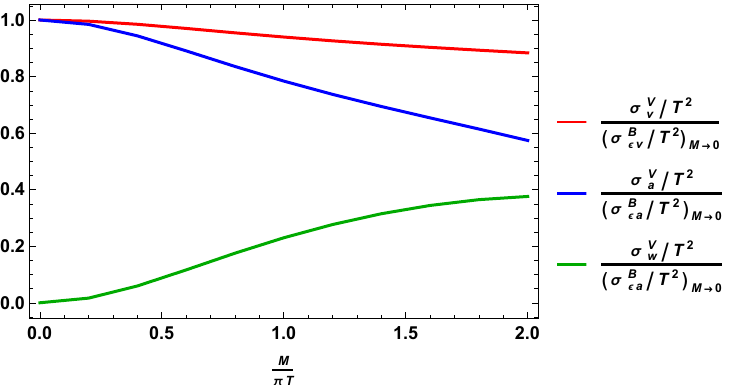}
		\caption{Plot of the chiral vortical conductivities $\sigma^V_s$, where $s \in \{ v, a, w \}$,  as a function of $M/(\pi T)$. See fig.~\ref{fig:comb1} for further details.}
		\label{fig:comb3} 
	\end{figure}

	In this subsection we will show the results for the behavior of the conductivities as a function of $M/(\pi T)$ for fixed non-vanishing chemical potentials.
	We will present below the results corresponding to fixing the chemical potentials as $\mu_v/(\pi T) = 0.1$,  $ \mu_a/(\pi T)=0.3$ and $\mu_w/(\pi T) = 0.2$.  We have computed the conductivities involving vector, axial, non-anomalous and energy currents, as a function of  $M/(\pi T)$. We show the plots of the conductivities normalized by their $M\to0$ values when the latter are finite. In cases where the conductivities vanish in this limit, we normalized them by using some related non-vanishing conductivity from the same sector. The results are given in figs.~\ref{fig:comb1}, ~\ref{fig:comb2} and~\ref{fig:comb3}. We can see in these figures that the conductivities inducing the axial current $\vec{J}_a$ are monotonically suppressed, while the conductivities inducing the non-anomalous current $\vec{J}_w$ become enhanced when $M/(\pi T)$ increases.
	
	\section{Discussion}
	\label{disc}
	In this work, we have studied the effect of explicit symmetry breaking in anomaly-induced transport by using a holographic framework that includes both pure gauge and mixed gauge-gravitational Chern-Simons terms. Our findings show that when symmetry is explicitly broken, anomalies can have a considerable impact not only on the anomalous currents but also affect the non-anomalous sector as well. 
	
	Our results reveal how symmetry breaking terms interact explicitly with anomaly contributions to modify the response to anomalous transport. One of the most interesting results of this work is that once the symmetry is broken by the scalar field $\phi$, even the non-anomalous current $\vec{J}_w$ shows response akin to the chiral vortical and chiral magnetic effect. While the present work has been carried out within a strongly coupled holographic framework, it would be interesting to study the analogous phenomenon in a weak coupling field theoretical setup~\cite{Landsteiner:2011cp}. Such a comparison would help to clarify which features are robust consequences of quantum anomalies and which are specific to the holographic regime. Another natural avenue is the exploration of phenomenological realizations. In particular, systems such as Weyl semi-metals typically come with a multiple of Weyl nodes (valleys) in the Brillouin zone~\cite{Basar:2013iaa,Landsteiner:2015lsa}. One can speculate that they feature not only axial symmetries by also some non-anomalous accidental valley symmetries and corresponding valley currents. It would be fascinating to see a similar non-anomalous chiral magnetic effect in such valley currents in certain phases in which parts of valley symmetries are broken.
	
	\section*{Acknowledgments}
	A.T. would like to thank his supervisor Shubhrangshu Ghosh for his valuable suggestions and help throughout the work. A.T. would also like to thank Inter-University Centre for Astronomy and Astrophysics (IUCAA), Pune for the hospitality during the Raman Memorial Conference held at Savitribai Phule Pune University (SPPU). A.T. acknowledges financial support by the Ministry of Tribal Affairs, Government of India, through the National Fellowship for Scheduled Tribes (NFST). The works of N.R. and E.M. are supported by the ``Proyectos de Investigaci\'on Precompetitivos'' Program of the Plan Propio de Investigaci\'on of the University of Granada under grant PP2025PP-18. The work of K.L. is partially supported by CEX2020-001007-S and PID2021-123017NB-I00, PID2021-127726NB-I00,  PID2024-156043NB-I00 funded by MCIN/AEI/10.13039/501100011033 and ERDF ``A way of making Europe''.  
	\bibliographystyle{unsrt}
	\bibliography{mybib}

\end{document}